# Classification of Chest Diseases using Wavelet Transforms and Transfer Learning

Ahmed Rasheed[1], [1]Muhammad Shahzad Younis, [2]Muhammad Bilal and [1]Maha Rasheed

[1] School of Electrical Engineering and Computer Science, National University of Sciences and Technology (NUST), Islamabad, Pakistan
`arasheed.msee17seecs@seecs.edu.pk`
[2] University of the West of England, Bristol, United Kingdom

**Abstract.** Chest X-ray scan is a most often used modality by radiologists to diagnose many chest related diseases in their initial stages. The proposed system aids the radiologists in making decision about the diseases found in the scans more efficiently. Our system combines the techniques of image processing for feature enhancement and deep learning for classification among diseases. We have used the ChestX-ray14 database in order to train our deep learning model on the 14 different labeled diseases found in it. The proposed research shows the significant improvement in the results by using wavelet transforms as pre-processing technique.

## 1    Introduction

Chest related diseases are considered to be in the leading causes of deaths. To diagnose these diseases, X-Ray is the most regularly used modality. Automatically classifying the diseases within X-Ray image scans of chest remains a tough task. X-Ray scan being a very cost effective exam for body inspection is performed frequently in medical checkups [1]. However, it has been quite a challenging task to clinically diagnose the chest diseases from the scans. Sometimes, this is believed to be more tough than the diagnosis using chest CT scan imaging. Some highly encouraging work has surfaced in the past, and also in recent researches. But the achievement of computer-aided detection and diagnosis (CAD) that is clinically relevant in practical medical stations on chest X-ray with all data settings is still very unfeasible, if not impossible with a dataset that has a handful of images for study and training [20]. Early detection of infected areas of chest and diseases can save lives of many. Computer aided systems are now of great interest to reduce the number of casualties by detecting it in early stages and helping out the radiologists in making important decisions.

The proposed system presents an effective network which utilizes the knowledge and techniques of image processing in the form of wavelet transforms of images and deep learning resulting in classification of infected regions found in the X-Ray scans of chests. The diseases are classified in the different classes based on the structure, shape and composition of infection present.

Training Convolutional Neural Network (CNN) model with deep connections from the scratch is not easy as a very large dataset and compute power is needed. One way to do the job is to use already trained deep models and tune them according to your needs. Transfer learning also does a similar job by removing and then adding small number of layers modified for the problem in study at the bottom of the network and then training only the newly added layers. This method is effective when you have smaller datasets. The dataset that this paper utilizes is the Chest X-Ray 14 database [19] that contains a total of 112,120 images from 30,805 patients. The dataset provides a considerably large as well as diverse repository for both training as well as testing. The approach to this paper is a comparison between the results retrieved via Transfer Learning in two cases; with wavelet transform and without it.

## 2    Background and Previous Work

Deep learning models can attain human like thinking and intelligence because of their complex structure which resembles the human neural system. There has been a significant amount of work done by scientists by using deep learning along with pre-processing and augmentation techniques for different problems. Some work has also been done in medical field to achieve state of art results for different problems having different datasets. While working on dataset comprising of images, the most common deep learning architecture used is CNN. Classification is widely done by using CNN architecture for image based data.

### 2.1    Wavelet Transform

Wavelet transform is a signal processing technique which can be used on images to extricate the important features within the image before feeding it to the classifying network. Through wavelet transform we get simultaneous localization in frequency and time domain [2]. Wavelet transforms have been used in multiple researches to help the doctors or radiologists in order to find the infected region from the X-Ray scans of patients [3-5]. It enhances the features in the imaging which then helps the neural networks to learn those features efficiently while training. It helps the network to learn the important features like the sharp edges, bright regions or other abnormalities found in the X-Ray scan [7].

The general equation representing the working of wavelet transform is provided below. Here in Eq. 1 $\psi$ represents the transforming function, $f(x)$ original signal, $k$ translation parameter and $j$ is scale parameter [8].

$$W_k^j = \int f(x)\psi\left(\frac{x}{2^j} - k\right) dx \tag{1}$$

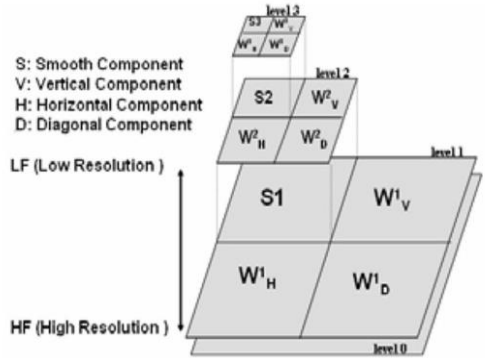

**Fig. 1.** 2D Wavelet transform layer representation up to 3 levels [8].

## 2.2    Augmentation

Augmentation is the process of generating new samples of data by changing its orientation. It improves the performance of the network and prevents it to over-fit for a specific problem [9]. Augmentation is mostly used when the total quantity of images in a dataset is not enough to train the network effectively [10]. In classification problems it also serves a great role in generalizing the network to perform well for all the classes and similar datasets [11].

## 2.3    Deep Neural Network (DNN)

DNN are widely used for the classification of medical images [12]. A lot of work has been done in medical field by using DNNs, detection of chest diseases has also been done before through DNN in order to obtain better results compared to conventional classification techniques [13].

**Convolutional Neural Network**  CNNs are composed of layers with number of convolutional filters which are learned by the network in such a way that they classify the image at the end of the network. CNN are used for processing the image based data as it learns the filters to distinguish between classes [12]. The end layer of the CNN works on the basis of probability as it outputs the probability of each class and then selects the class having the highest probability for an image.

**Transfer Learning**   Also known as "Transfer-ability" is the use of network trained on a different dataset and problem for your own dataset and problem. In transfer learning, a pre-trained network is obtained and its last few layers are replaced in order to amend its architecture for the classification of new classes, then the filters of first few layers of network are kept frozen, remaining layers in the network are further trained on the newer dataset. Due to the limited amount of data available, associated with medical imaging, transfer learning comes in handy instead of training from scratch [14,15].

Few known pre-trained networks are ResNet, VGG, GoogleNet and AlexNet. All these networks have been trained on the ImageNet database [16] designed to classify images with 1000 classes. To use them for the proposed problem first few starting layers are kept frozen and newly added last layers are trained again in order to work for the desired classes. Learning rate is set high for the new layers added. The network above remains unchanged and the new layers are further trained for the newer dataset.

## 3      Dataset

Neural networks generally require a large number sample images for the training purpose. The dataset used for this research is ChestX-ray14 database. This dataset is obtained from the online database formulated by the U.S. National Institute of Health Clinical Center, through their clinical PACS. This database has 60% scans from all frontal chest X-ray scans done within the hospitals [19].

Chest X Ray 14 database consists of total 112,120 images of frontal X-Ray scans collected from 30,805 patients having fourteen different diseases. Dataset is obtained through natural language processing of their associated radio-logical reports. There are fourteen common thoracic pathologies found in the chest X-ray scans. Table 1 shows the names and distribution of all these fourteen classes based on the total images existing in each class.

**Table 1.** Class distribution of database

| Classes | Number of Images |
|---------|------------------|
| Infiltration | 25366 |
| Effusion | 18974 |
| Atelectasis | 16057 |
| Nodule | 8409 |
| Mass | 8269 |
| Consolidation | 7177 |
| Pneumothorax | 7134 |
| Pleural Thickening | 5172 |
| Cardiomegaly | 3906 |
| Emphysema | 3586 |
| Edema | 3443 |
| Fibrosis | 2211 |
| Pneumonia | 2092 |
| Hernia | 284 |
| Total | 1121120 |

For training and testing purposes the dataset is further divided in to three sub parts. Training, validation and test data at 75%, 15% and 15% respectively. This splitting of dataset is totally random to ensure unbiased network training.

## 4      Methodology

The flow diagram in the Figure 2 gives the overall idea of the methodology used.

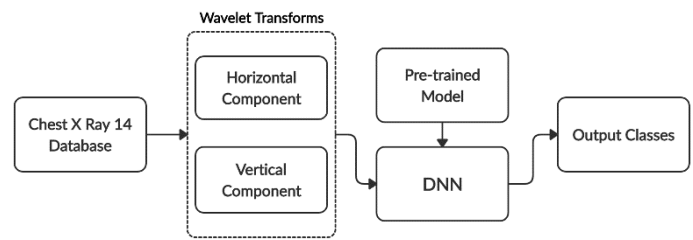

**Fig. 2.** Proposed system workflow diagram.

### 4.1      Wavelet Transforms

Wavelet transform of the X-Ray scans of complete dataset are done as image pre-processing in the proposed system. They help the network by enhancing the features in the images. Two output images are obtained from this process, one in vertical and other in horizontal directions of image. Sample images are shown in Figure 3. Images are resized and then fed to the DNN for classification

### 4.2      Deep Neural Network (DNN)

After the pre-processing of images which is wavelet transform and resizing, the images are then fed to the DNN for classification purpose. Rather than training a whole network from scratch we have used transfer learning techniques on a pre-trained network.

**Pre-trained CNN**  An already trained networks is more useful instead of building a whole new CNN architecture from scratch and for this purpose we have used ResNet50 architecture which was introduced in 2015 [17]. It is one of the state of art network trained for the ImageNet database for classification task [16].

**Augmentation**  In order to generalize the train network and avoid the over-fitting problem, we have done augmentation by doing rotation, translation and scaling with random parameters.

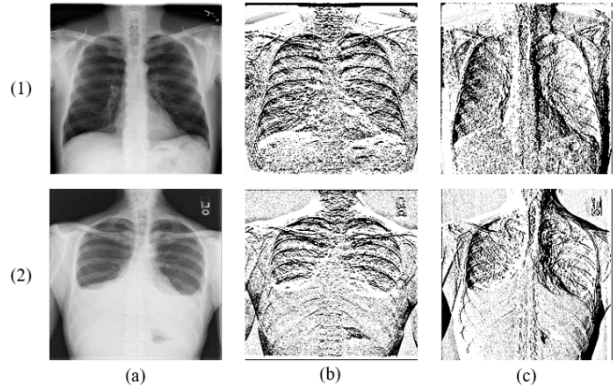

**Fig. 3.** Samples obtained from wavelet transform for two diseases (1) Infiltration, (2) Effusion while (a) is original and (b)(c) are vertical and horizontal component of image respectively.

**Transfer Learning**    In our proposed system we have transferred the knowledge of already trained network on ImageNet database to our domain which is classification of diseases in the ChestX-ray14 database by freezing the first 20 layers of pre-trained network (ResNet50) and replacing its final layers to classify our desired classes. Transfer learning comes in handy when the data on which the system is to be trained is a small database [18].

**Training and Validation**    The dataset is split onto three parts, training, validation and test dataset. The training dataset has been used to train the network by updating its weights for each layer added while the validation dataset serves as a feedback to the network after every few iterations and helps the system to improve and avoid overfitting. Table 2 lists the hyper-parameters used for the training of networks.

**Table 2.** Hyper-parameters of Network

| Parameters | Values |
|---|---|
| Total Epochs | 15 |
| Batch Size | 20 |
| Initial Learning Rate | 3e-4 |
| Activation Function | ReLu |
| Optimizer | SGDM |

## 5    Results and Discussion

We performed two kind of tests. First DNN was trained using transfer learning on the original dataset with 14 classes. Then the training of network was performed on the images obtained after using wavelet transform as pre-processing technique. The evaluation metric for the results is Receiver Operating Characteristic (ROC) curve. Curves for both the tests are shown in the Figure 4 for the three major classes from the dataset.

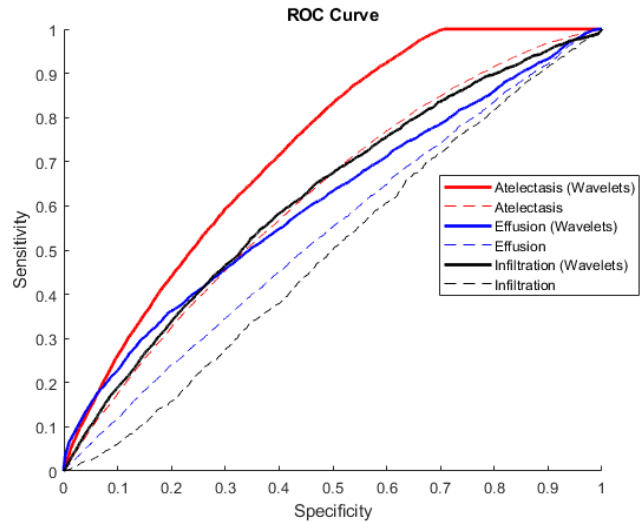

**Fig. 4.** ROC for three major classes Atelectasis, Effusion and Infiltration.

The solid lines represent the ROC for the network trained on wavelet transform of images while dotted lines represents the network trained on original images. We can see from Figure 4 that taking wavelet transform of X-Ray scan significantly improves the performance of the classification network. As for future work, more complex DNN architectures can give much better results when trained using wavelet transforms on high end compute resources.